\documentclass[aps,pra,twocolumn,groupedaddress,showpacs]{revtex4}

\usepackage{graphicx,amsmath,hyperref}
\usepackage[latin1]{inputenc} 
\usepackage{amsfonts}
\usepackage{amssymb}
\usepackage{color}
\usepackage{graphicx}

\newcommand{\ket}[1]{|#1\rangle} 
\newcommand{\bra}[1]{\langle#1|} 

\begin{document}

\title{Adiabatic Computation - A Toy Model}

\author{Pedro Ribeiro}
\email[]{pedro@lptmc.jussieu.fr}

\author{Rémy Mosseri}
\email[]{mosseri@ccr.jussieu.fr}

\affiliation{Laboratoire de Physique Théorique de la Matière Condensée,
             Université Pierre et Marie Curie, Place Jussieu, 
             75252 Paris Cedex 05, France}

\date{\today}



\begin{abstract}
We discuss a toy model for adiabatic quantum computation which displays some phenomenological properties expected in more realistic implementations. This model has two free parameters: the adiabatic evolution parameter $s$ and the $\alpha$ parameter which emulates many-variables constrains in the classical computational problem. The proposed model presents, in the $s-\alpha$ plane, a line of first order quantum phase transition that ends at a second order point. The relation between computation complexity and the occurrence of quantum phase transitions is discussed. We analyze the behavior of the ground and first excited states near the quantum phase transition, the gap and the entanglement content of the ground state.
\end{abstract}



\maketitle

\section{ Introduction}

Adiabatic Quantum Computation (AQC) has been proposed in 2001 by Farhi et al. \cite{FarGolGutSip00} as an alternative way to perform quantum computation. It is based on the quantum adiabatic theorem \cite{Mes76} which states that a system initially in its ground state, and subject to a sufficiently slow varying time dependent Hamiltonian will remain in the ground state as long as there is no energy level crossing in the course of the evolution. The protocol proposed by the authors consists in passing from a initial Hamiltonian $H_0$ with a well known and easy to prepare ground state to the ground state of a final Hamiltonian (or problem Hamiltonian) $H_P$ which encodes the answer to a given classical problem. This approach has been proved to be equivalent to the standard quantum gate model of computation \cite{DamMosVaz01},\cite{AhaDamKemLanLloReg04}. The protocol of adiabatic computation amounts therefore to a controlled path in the space of hermitian operators acting on an Hilbert space of a physical system ($H(s),\ s \in [0,1] $), which is the tensor product of $n$ two level system (quantum bits). The path is usually taken to be the linear interpolation between $H_0$ and $H_P$ , resulting in a total Hamiltonian:
\begin{equation} \label{eq:H_s}
H(s) = (1-s) H_0 + s H_P .
\end{equation} 
The time $T$ taken to perform the computation is such that the adiabatic theorem applies, with the probability of passing to an excited state remaining limited \cite{Mes76}, which roughly translates to:
\begin{equation} \label{eq:time_gap}
T >> \Delta_{min}^{-2},
\end{equation} 
where $\Delta_{min}$ is the minimum value of the energy difference between the ground and the first excited states taken along the evolution. Therefore the time scaling with $n$ (and so the computational efficiency) will be mainly determined by the behavior of the energy gap between the two lowest energy states, and is rather long whenever this scaling is exponentially small. Up to now only exponential decreasing gaps were proved to exist for some adiabatic protocols trying to solve NP-complete problems (\cite{DamMosVaz01, FarGolGutNag05, Rei04, wei-2006-, ZniHor05}) and some results, namely for the 3-SAT problem are rather inconclusive (\cite{FarGolGutSip00, FarGol00, FarGolGutLapLunPre01}). For the unsorted data base search a AQC algorithm was found that reproduce the gain of the Grover's algorithm for standard quantum computation  \cite{RolCer03}. Even if no new improvements in algorithm design were obtained up to now using this approach, an advantage pointed out by \cite{ChiFarPre02} is the robustness of this protocol against quantum errors.

\section{ The Adiabatic Algorithm}

\subsection{General Remarks}

Let us first discuss the general properties of the Hamiltonians considered in this paper. We scale all Hamiltonians such that they have a bounded spectrum. Even if for a typical physical system one expects the energy to be proportional to $n$, this is just a linear scaling and can be later taken into account in the total computational time for a realistic physical system. 
\\
\\
Given a classical computational problem one should first map its solution(s) onto a ground state of an Hamiltonian denoted $H_P$. We assume that the final measurement (output of the computation) is performed in the computational basis, and so, that the final Hamiltonian is diagonal in this basis.
Once $H_P$ is defined, one can ask for an optimal initial Hamiltonian $H_0$ and an optimal path $H(s)$ in the parameters space (that maximizes the energy gap).
\\
\\
In this paper we will only consider Hamiltonian paths of the simple form (\ref{eq:H_s}). This choice is motivated by the following arguments. First it is clear that any Hamiltonian path of interest will be such that $[ \partial_s H(s), H(s)] \neq 0$. Indeed suppose that for a path $H(s)$ there exist an interval of values of $s \in [s_0, s_1]$ such that we have $[ \partial_s H(s), H(s)]=0$, implying that Hamiltonians within that interval commute with each others. In this situation two cases may occur: either there is one or more level crossings between the ground state and the first excited state and, in that case, the adiabatic condition is no longer valid; or the ground state experiences no level crossing (so its correspondent eigenvectors remains the same) and the evolution amounts to a rescaling of the energies which could be performed "instantaneously" without breaking the adiabatic condition, because all the non-diagonal matrix elements of $\partial_s H(s)$ vanish. In this case the evolution between $s_0$ and $s_1$ needs not to be done adiabatically. In the following we suppose that all Hamiltonian paths do not have such ``commuting`` intervals and so we end up with a path such that $[ \partial_s H(s), H(s)]\neq0, \ \forall s\in[0,1]$, which is the nontrivial part of the protocol. 
\\
\\
It should be clear that in most interesting cases the system will undergo a Quantum Phase Transitions (QPT) along the Hamiltonian path. Indeed the complexity of a classical computational problem translates into an increasing of the computational time with n. In the framework of the adiabatic theorem, this implies that there exists at least one value of s such that a gap closes (with increasing n) during the adiabatic evolution. Passing a vanishing gap corresponds to QPT in physical terms. In the vicinity of a QPT the system is described by its universality class which depends on the relevant couplings of the Hamiltonian. The family of Hamiltonians with the same relevant couplings presents the same behavior at the QPT and so share the same complexity when considered as adiabatic algorithms.
\\
\\
The relation with QPT suggests that a given algorithm (path) should not fundamentally depend on "small details" of this path but rather on some relevant features near the QPT. This is why the simple interpolation Hamiltonian path (\ref{eq:H_s}) can be chosen. Given $H_P$ the choice of the path is then reduced to the choice of $H_0$

\subsection{ The Algorithm}

The output of the adiabatic computation is the ground state of the final Hamiltonian $H_P$. We build such a Hamiltonian by attributing to each possible classical configuration $x_i$, between the $ N = 2^n$ possible ones, a real value $\varepsilon_i$ (energy) that measures how well the problem is satisfied by the string of bits $x_i$. If $x_i$ is a solution of the problem we set $\varepsilon_i=0$, otherwise $\varepsilon_i$ takes a non zero value, usually based on the problem statement (for example for the 3-SAT problem, $\varepsilon_i$ is the number of clauses violated by the string $x_i$). The problem Hamiltonian reads
\begin{eqnarray}
H(s=1) = H_P = \sum_i  \varepsilon_i \ket{i} \bra{i},
\end{eqnarray} 
where $\ket{i}$ runs over the $N=2^n$ states in the computation basis which we take to be the tensor product of the individual eigenstates of $\sigma_z$ for a two level system. If the physical system remains in its ground state, the final output corresponds to the minimization of the energy $\varepsilon_i$ as a function of $i$. For sake of simplicity we take $H_0$ to be diagonal in the $x$ basis. Concerning the $s$ parameter, we assume through this paper that it is a simple affine function of the time such that $s(0)=0$ and $s(T)=1$. Note that some authors have proposed to speed up the adiabatic evolution far from the QPT point leading to an efficient gain for the total computational time \cite{RolCer03}.

\section{ The Model}

In the previously published papers on AQC two different prescriptions of the initial and final Hamiltonians were used. The most common is an additive Hamiltonian made of interaction terms involving few qubits (pairs and triplets). Indeed this type of problems is usually given by a set of local constrains (concerning few variables e.g. 3 for the 3-SAT problem). This additive prescription can also be used for $H_0$ providing that in that case the ground state can be easily prepared. The other type of Hamiltonians are projector-like ($H^2\sim H$)  \cite{FarGolGutNag05, ZniHor05, RolCer03}. A $H_P$ of this type corresponds to an oracle-based problem which has two possible values of the energy: a (possibly degenerate) ground state energy whose states are the problem solutions and an excited energy for non solutions.
\\
\\
We are interested in studying different types of adiabatic evolutions which differ in terms of the gap scaling. For that purpose we chose a fixed initial Hamiltonian $H_0$ and we study a range of $H_P$ with different gap scaling properties. Since $H_0$ should be easy to implement and diagonal in the $x$ direction, the most natural choice is a simple (normalized) magnetic field interaction along the $x$ direction:
\begin{eqnarray}
H_0 =  \frac{\mathbb{I}}{2} - \frac{1}{n} S_x,
\end{eqnarray}
where $S_x = \frac{1}{2} \sum_{k=1}^{n}  \sigma_x^{(k)}$.
We aim to present a simplified model of adiabatic computation by using an Hamiltonian $H(s)$ which is solvable, while displaying some of the features which will eventually be found in the more realistic case, namely the spectrum diagonal in the $z$ basis for $H(s=1)$, and a quantum phase transition at some intermediate value $s_c$. We choose a hermitian operator $h_p$ with $k$ spin interaction terms of the form of the following tensor product
\begin{eqnarray}
h_p & = &  \otimes_{i=1}^n(\mathbb{I}^{i} + p \sigma^i_z),
\end{eqnarray}
where $i$ denotes the qubit. It is clearly a sum of $k$-spin interaction terms ($k$ ranging from $1$ to $n$), whose strength depends on $p$. Introducing the total spin $S_z=\frac{1}{2}\sum_{i=1}^{n}\sigma^i_z $, and using the identity $e^{\alpha \sigma_z} = \cosh(\alpha) + \sinh(\alpha) \sigma_z$, $h_p$ can be written as 
\begin{eqnarray}
h_p & = & (1-p^2)^{-\frac{n}{2}} e^{2 \tanh^{-1}(p) S_z}.
\end{eqnarray}
Finally, introducing $\alpha = n \tanh^{-1}(p)$, we rescale $h_p$ into $H_P$ in the following form
\begin{eqnarray}\label{HP}
H(1) = H_P(\alpha) & = &   \frac{ e^{ \alpha \mathbb{I}} - e^{2 \frac{\alpha}{n} S_z } }{2 \sinh \left(  \alpha \right) }.
\end{eqnarray}
Note that the $n$ factor in the definition of $\alpha$ is introduced to obtain an Hamiltonian $H_P$ such that $n H_P$ is an extensive operator.
\\
\\
The ground state of this Hamiltonian is $\ket{w}= \otimes_{k=1}^n \ket{0}$ which has zero energy. The $m$th excited states correspond to a state with $m$ $1$s  having a binomial degeneracy $\frac{n!}{(n-m)! n!}$ for any finite value of $\alpha$ and whose energy depends on $\alpha$ (see below).
For $\alpha\rightarrow\infty$ the Hamiltonian is proportional to a projection operator:
\begin{eqnarray}
H_P(\infty) =    \left(\mathbb{I} -  \ket{w }\bra{ w}\right) ,
\end{eqnarray}
and all the excited states have energy equal to one. The final Hamiltonian obtained this way is the Grover-like unsorted data base searching considered in \cite{ZniHor05, DamMosVaz01} and \cite{RolCer03}. The limit $\alpha \rightarrow 0$ decouples the qubits and $H(s)$ can be written as a sum of independent single qubit Hamiltonians:
\begin{eqnarray}
H(s,\alpha \rightarrow 0) =  \frac{1}{n} \sum_{k=1}^{n} (1-s) \frac{\mathbb{I}-\sigma_x^{(k)}}{2} + s \frac{\mathbb{I}-\sigma_z^{(k)}}{2},
\end{eqnarray} 
which corresponds to a trivially separable problem that can be solved by parallelizing single qubit problems.
\\
\\
Note that, although the above $H_p$ ground state has a particular simple form, the properties described below would apply to any Hamiltonian obtained from $H(s)$ under unitary transformations. Suppose a given problem has a solution corresponding to the ground state $\ket{w({J})}=\otimes_{k=1}^n \ket{J_k}$, where $J_k\in \lbrace0,1\rbrace$, and the same energies and degeneracies as the $H_P$ described above. In this case by performing the unitary transformation:
\begin{eqnarray}
U =  \bigotimes_k (\sigma_x^{(k)})^{J_k}
\end{eqnarray} 
to $H(s)$, $H_0$ remains invariant and the ground state of $H_P$ is changed to $\ket{w({J})}$. Since $U$ is unitary the gap scaling nature and the energy spectrum stay invariant. So, upon studying the particular form of $H_P$ (\ref{HP}), we have access to the gap behavior of problems which can have the whole set of possible $2^n$ ground states i.e. solutions.
The advantage of $H_P(\alpha)$ as a representant of this class is that it is symmetric under qubit permutation and so it can be written as a function of the total spin $S_a = \frac{1}{2} \sum_k \sigma_a^{(k)}$ for $a=x,y,z$. Moreover since $H_0$ and $H_P(\alpha)$ commute with $S^2=\sum_a S_a^2$ and the ground state of $H_P$ has maximal value for the total spin $j=\frac{n}{2}$, the whole evolution will take place in this maximal spin sector spanned by the Dicke basis: $\lbrace \ket{ \frac{n}{2},i}\rbrace_{i=-\frac{n}{2}}^{\frac{n}{2}} $, where the two quantum numbers stand for the value of the total spin ($j=\frac{n}{2}$) and the spin projection along the $z$ axes. All the other spin sectors can be disregarded because they are not coupled by the adiabatic evolution \cite{Mes76}.

\subsection{Density of states}

\begin{figure}[ht]
\includegraphics[width=0.5\textwidth]{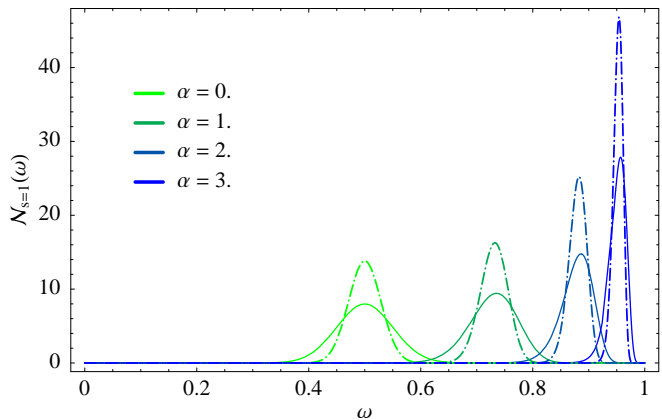}
	\caption{\label{fig:densityofenergy}  Density of states as a function of the energy for $s=1$ for different values of $\alpha$. The solid lines correspond $n=100$ and dashed lines to $n=300$, plotted for $\alpha \in \lbrace 0, 1, 2, 3\rbrace$.   }	
\end{figure}

As explained above the complexity of the adiabatic evolution is related to what happens near the QPT, it is nevertheless interesting to describe the spectral properties of $H(s)$ for the extremal values of $s$ (equal to zero and one).
The density of states as a function of the energy for $s=0$ and $s=1$ can be given analytically for large $n$.
For $s=0$, considering the binomial degeneracy of each level, the density of states writes:
\begin{eqnarray} 
 \mathcal{N}_{s=0} (\omega)  d\omega &=&   \frac{2 n}{\pi} e^{-2n(\omega - \frac{1}{2})^2}  d\omega, 
\end{eqnarray}
where $\mathcal{N}_{s} (\omega)  d\omega$ is the total number of levels between $\omega$ and $\omega+d\omega$.
For the case $s=1$ one starts by remarking that the energies $\omega_m$ in the maximal spin sector are given by:
\begin{eqnarray} 
\omega_m (s=1) = \frac{e^\alpha - e^{ \alpha (2\frac{m}{n}-1) }}{ 2 \sinh \alpha},
\end{eqnarray}
where $m$ is the level labeling $m = 0, ..., n$. For large $n$  one can invert this relation and obtain the energy density for the maximal spin sector:
\begin{eqnarray} 
 \mathcal{N}_{s=1,S=\frac{n}{2}} (\omega)  d\omega= \frac{1}{\alpha \left( 1 - 2 \omega + \coth \alpha \right) } d\omega,
\end{eqnarray}
To get the density of states as a function of the energy for the total spectrum one has to consider the binomial degeneracy of each level:
\begin{eqnarray} 
 \mathcal{N}_{s=1} (\omega)  d\omega &=&    \frac{2 n}{\pi} \frac{e^{ -2n(\frac{\alpha + \ln(e^\alpha - 2 \omega \sinh \alpha )}{2 \alpha} - \frac{1}{2})^2}}{\alpha \left( 1 - 2 \omega + \coth \alpha \right) } d\omega.
\end{eqnarray}
The behavior of the density of states at $s=1$ as a function of $n$ and $\alpha$ is shown in Fig. \ref{fig:densityofenergy}. Remark that for $\alpha=0$ one recovers the Gaussian centered at $\omega=\frac{1}{2}$ resulting from the binomial degeneracy of the separable problem. For increasing $\alpha$, the density of states is more peaked toward the value  $\omega=1$ which characterizes the projector-like Hamiltonian.
%

\section{ Mean Field Approach}

Since each qubit interacts in an equivalent way with all the other qubits, we expect that, in the thermodynamic limit, a mean field approach will give access to the exact ground state energy and signal quantum phase transitions whenever they occur. This is done by injecting the separable ansatz state:
\begin{equation}
\ket{\Psi \left( \theta , \phi \right)  }= \otimes_{i=1}^n \left( \cos(\frac{\theta}{2} ) \ket{0} + \sin (\frac{\theta}{2}) e^{i \phi } \ket{1} \right),
\end{equation}
and minimizing the energy in order to determine the free parameters $\theta$ and $\phi$.
Doing so we can see that a first order quantum phase transition occurs for some value of $s=s_c(\alpha) ( \approx 2.598 )$ providing that $\alpha > \frac{3 \sqrt{3}}{2}$.
\begin{figure}[ht]
\includegraphics[width=0.5\textwidth]{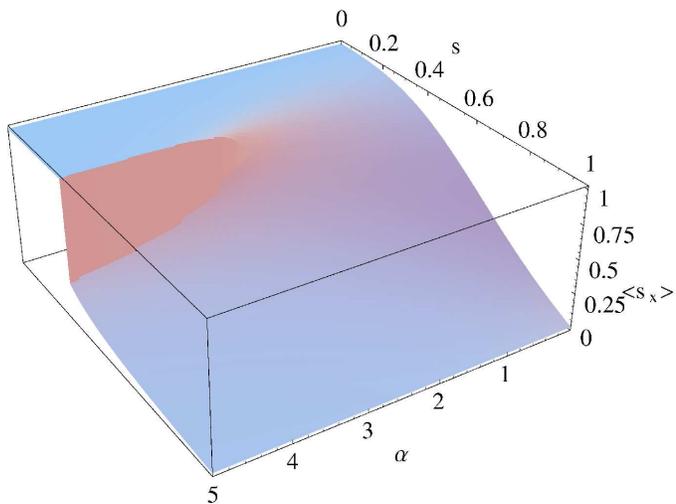}
\caption{\label{fig:Sx} Mean value of $s_x=\frac{S_x}{n}$ plotted in the  $\alpha-s$ plane.}	
\end{figure}
Fig.\ref{fig:Sx} shows the mean value of the observable $s_x=\frac{S_x}{n}$ which presents a discontinuity along the first order quantum phase transition line in the $\alpha-s$ plane. This line ends with a second order point $(\alpha, s)=(\frac{3 \sqrt{3}}{2}, \frac{2}{2+3\sqrt{6} e^\frac{3}{2} \sinh\left( \frac{3 \sqrt{3}}{2} \right) ^{-1}} )$ where the values of observables are continuous non analytic functions of $\alpha$ and $s$. For $\alpha > \frac{3 \sqrt{3}}{2}$, a discontinuity of $\langle S_x \rangle$ is related to the abrupt change of the ground state component with respect to the fully polarized states in the $x$ ($\ket{\Rightarrow}=\otimes_{i=1}^n\left( \frac{1}{\sqrt{2}} (\ket{0}+\ket{1})\right)  $) and $z$ directions ($\ket{\Uparrow}=\otimes_{i=1}^n\ket{0} $).

\section{  Numerical Analysis}

\subsection{ Energy Spectrum}
\begin{figure}[ht]
\includegraphics[width=0.5\textwidth]{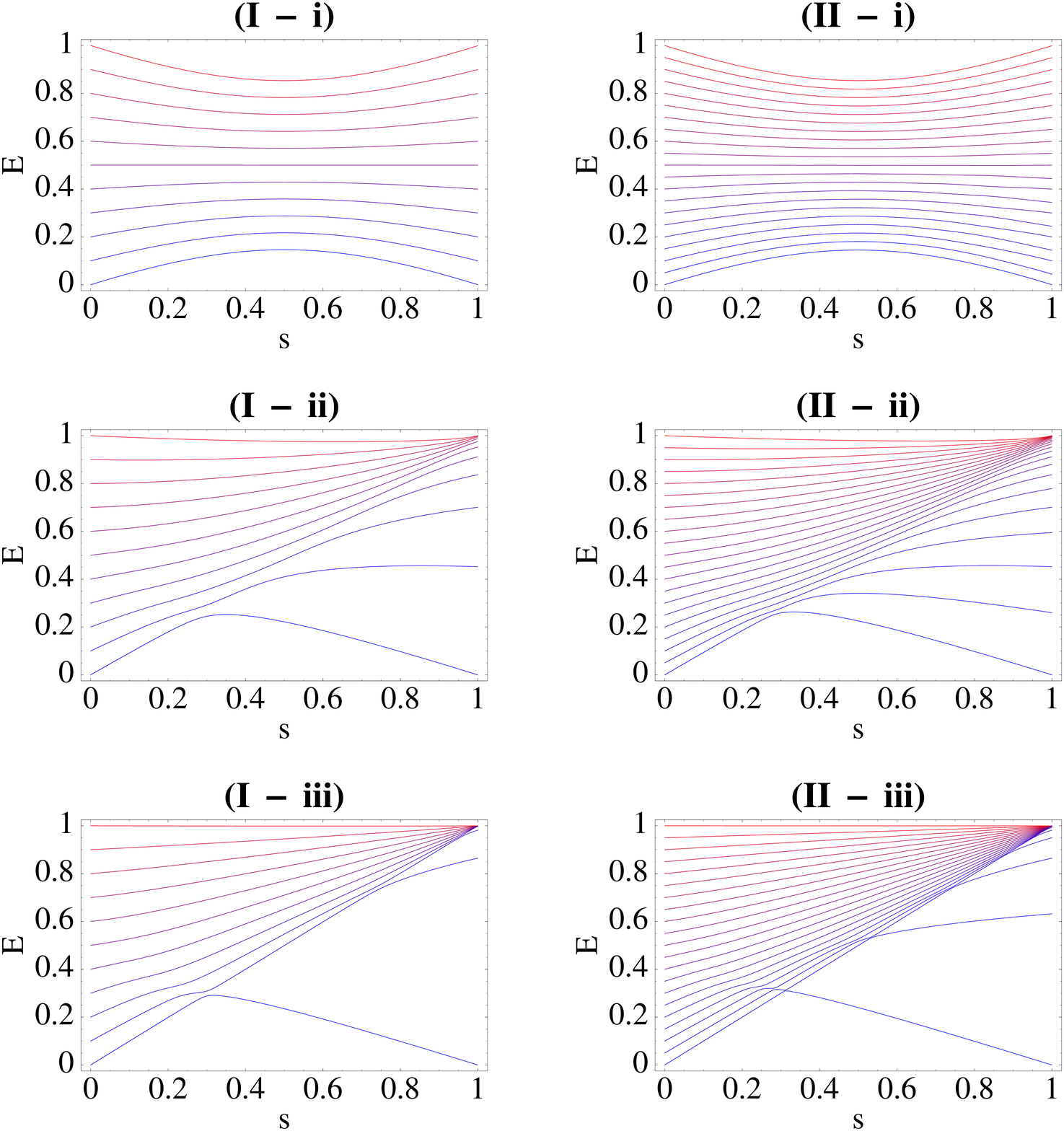}
\caption{\label{fig:energies} Energy levels as a function of $s$ computed for different values of $n$ and $\alpha$. The series $I$ presents the energy levels for $n = 10$ and $\alpha \in \lbrace 0,3,10\rbrace$ respectively for $\lbrace i,ii,iii\rbrace$. The series $II$ presents the energy levels for $n = 20$, the values of $\alpha$ are labeled as before. }
\end{figure}

As predicted by the mean field approach the value of the ground state energy tends to a maximum around $s=s_c(\alpha)$ and approaches the mean field ground state energy for increasing $n$. At the critical point there is an energy level anti-crossing and the gap vanishes as $n$ increases. Fig. \ref{fig:energies} shows the energy spectrum (in the symmetric sector) as a function of $s$ for different values of the $\alpha  $ parameter and $n$. 

\subsection{ Analysis of the Two Less Energetic States}

\begin{figure}[ht]
\includegraphics[width=0.5\textwidth]{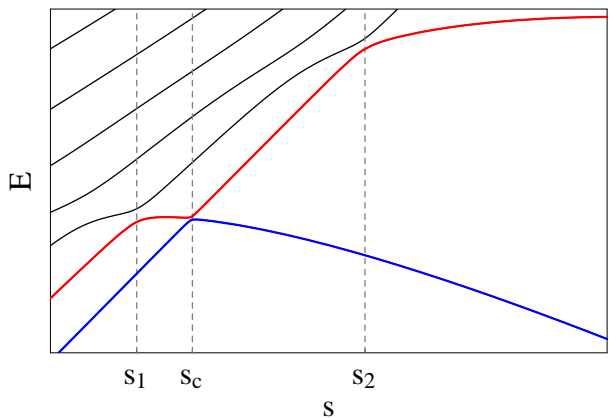}
\caption{\label{fig:Regions1} Different regions where the ground state (blue) and the first excited state (red) undergo level anti-crossings, plotted for $n=30$ and $\alpha=5$.}	
\end{figure}

\begin{figure}[ht]
\includegraphics[width=0.5\textwidth]{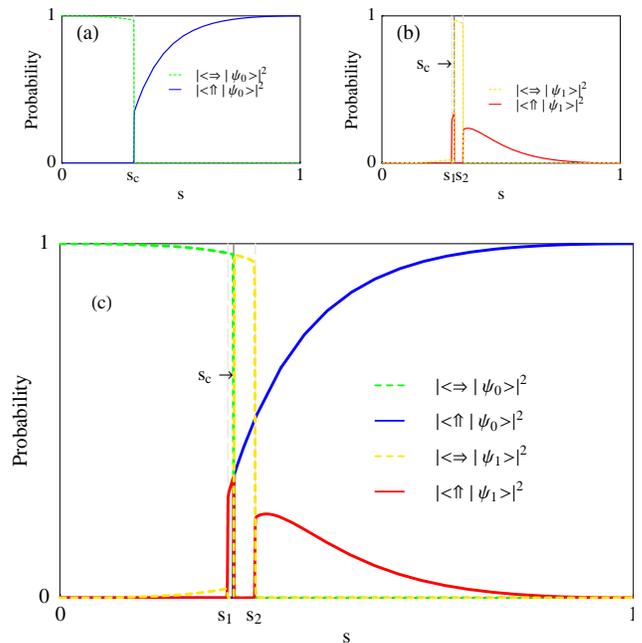}
\caption{\label{fig:Regions2}  Square of the modulus of the inner product of the two lowest levels with the states $\ket{\Rightarrow}$ and $\ket{\Uparrow}$ which are the ground state of $H(s)$ for $s=0$ and $s=1$ respectively. Here plotted for $n=50$ and $\alpha=5$. (a) and (b) present respectively the evolution of the ground and first excited states projections along the Hamiltonian path. (c) combines (a) and (b), and clearly displays the exchange between these two states at $s=s_c$.  }	
\end{figure}

\begin{figure}[ht]
\includegraphics[width=0.4\textwidth]{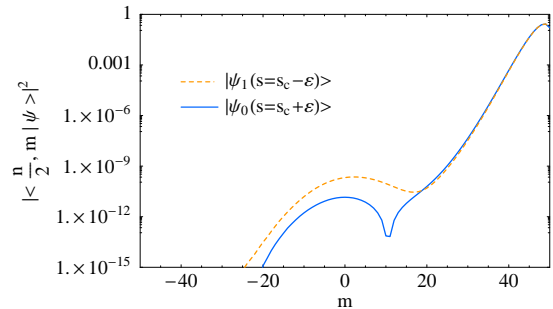}
\caption{\label{fig:Regions3} Square modulus of the projections of the ground state immediately after the anti-crossing and of the first excited state immediately before the anti-crossing along the Dicke basis, plotted for $n=100$ and $\alpha = 5$. Note that the a logarithmic scale is used.}
\end{figure}

For the cases where a phase transition exists ($\alpha>\alpha_c=\frac{3\sqrt{3}}{2}$) we now analyze the behavior of the two less energetic states, in particular their projections along $\ket{\Rightarrow}$ and $\ket{\Uparrow}$. Fig.\ref{fig:Regions1} presents a zoom of the ground state anti-crossing (at $s=s_c$) with the first excited state anteceded by some anti-crossings between more energetic states. As $n$ increases this cascade of anti-crossings gets closer to the QPT point.
Fig.\ref{fig:Regions2} shows the projections of the ground state and the first excited state along the fully polarized states $\ket{\Rightarrow}$ and $\ket{\Uparrow}$. We observe four different regions limited by the values of $s$ where  $|\left\langle  \Rightarrow | \psi_{0/1} \right\rangle |^2$ or $|\left\langle  \Uparrow | \psi_{0/1} \right\rangle |^2$ change abruptly.
In the region $0 > s > s_1$ there are several level anti-crossings  between excited states but they do not affect significantly the first two states of the spectrum (Fig.\ref{fig:Regions1}); we have $\ket{\psi_0} \simeq \ket{\Rightarrow}$ (Fig.\ref{fig:Regions2}) and $\ket{\psi_1} \simeq \ket{\frac{n}{2},\frac{n-1}{2}}_x$ up to a very good approximation. At $s_1$ the first excited state suffers an anti-crossing with the second one and its projection along $\ket{\Uparrow}$ increases drastically but remains different from one.
Fig.\ref{fig:Regions3} shows the projections onto the Dicke basis of the ground state immediately after the anti-crossing and of the first excited state immediately before the anti-crossing ($s=s_c$). At $s=s_c$ the transfer of components between the ground state and the first excited state is clearly manifested.
For $s>s_c$ the ground state increases slowly its projection along $\ket{\Rightarrow}$ and at $s_2$ the first excited state experiments another anti-crossing (Fig.\ref{fig:Regions1}), increasing drastically his projection along $\ket{j=\frac{n}{2},m=\frac{n}{2}-1}$. 
For $s>s_2$ there are no anti-crossings and the states increase slowly their projection along the ground and first-excited states: $  \ket{\Uparrow} $ and $\ket{j=\frac{n}{2},m=\frac{n}{2}-1}$.

\section{Beyond the Mean Field Approximation}

In order to characterize further the system let us now look at the gap scaling (with n) near the QPT and the ground state entanglement content as measured by the concurrence \cite{wootters-1998-80}. To study the quantum fluctuations around the mean field solution we can for instance apply a method described in \cite{dusuel-2005-71} which uses the  Holstein-Primakoff mapping from a given spin sector characterized by the value of $S^2$ and the algebra of boson operators, obtaining a interacting boson Hamiltonian which can be expanded in powers of $\frac{1}{n}$. Using this method one is able to compute the ground state concurrence in the  $\alpha-s$ plane and to predict the gap behavior based on the universality class of the model.

\subsection{Concurrence}

\begin{figure}[ht]
\includegraphics[width=0.5\textwidth]{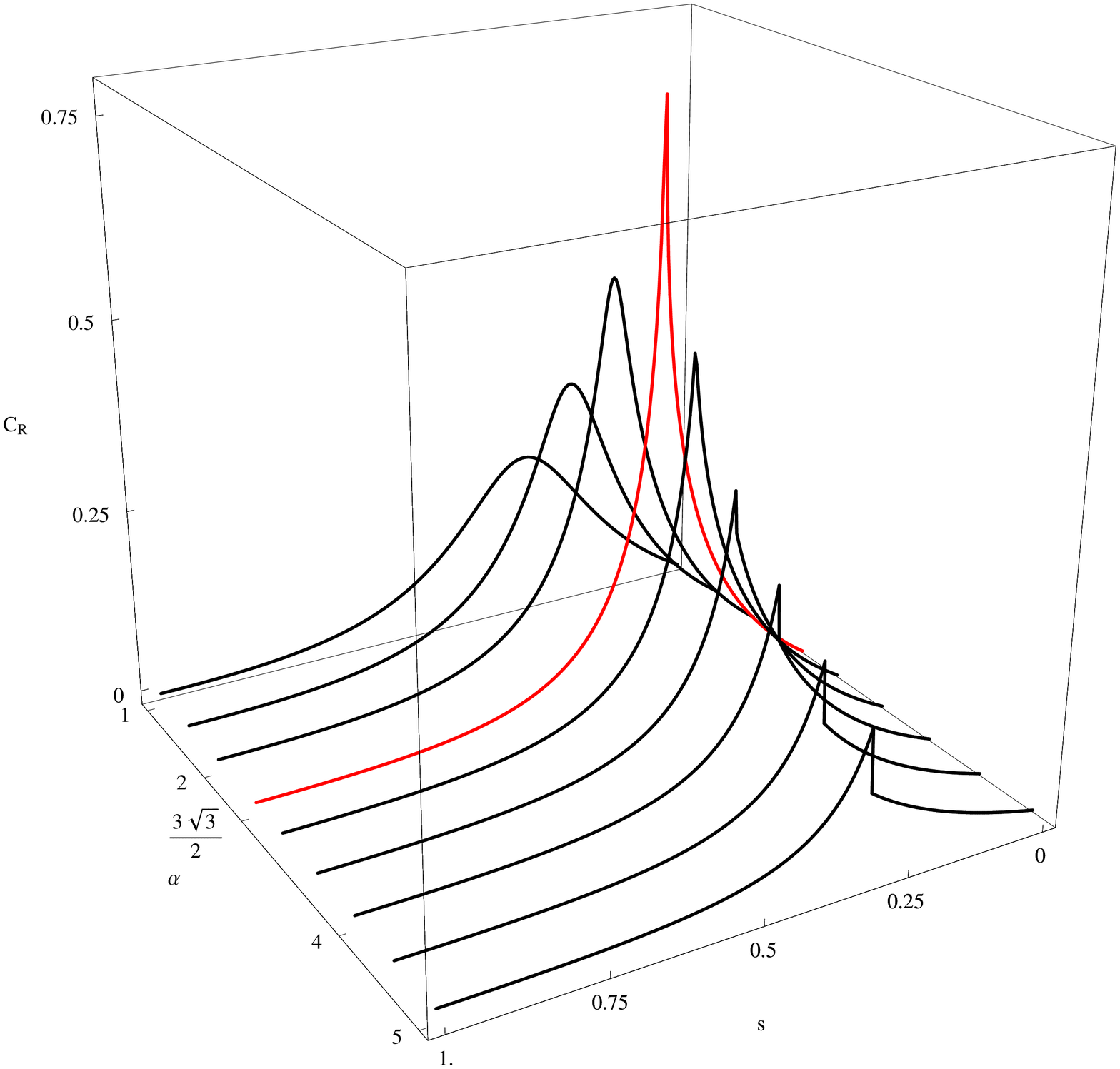}
\caption{\label{fig:concurrence} Reduced concurrence in the thermodynamic limit obtained, by tracing over $n-2$ arbitrary qubits, as a function of $\alpha$ and $s$. For $\alpha>\alpha_c $ the concurrence presents a discontinuity which increases with $\alpha$.}	
\end{figure}

We measure the entanglement contend of the ground state by computing the concurrence of the ($n-2$)-qubit traced density matrix.
The entanglement is encoded in the finite size corrections so the quantity to study is the rescaled concurrence $C_R = (N-1) C$ which is usually non trivial in the thermodynamic limit near a QPT. This quantity can be computed, for the symmetric spin sector, as a function of the mean values of the spin operators $S_i, S_i^2$, $i=x,y,z$, \cite{wang-2002-18}. 
In the present case the real nature of the density matrix leads to the simple expression: $C_R=1-\frac{4 \langle S_y^2 \rangle }{n}$: see \cite{vidal-2006-}.
 Fig.\ref{fig:concurrence} displays the concurrence computed in the $\alpha-s$ plane. Note that along the first order line the concurrence is discontinuous in $s$. At the second order transition point this quantity presents a cusp like form as in the simple LGM model \cite{vidal-2004-69}. The entanglement entropy can also be computed for this model using the method discussed in \cite{barthel-2006-} and also displays a discontinuity along the first order line \cite{pedro-2006-}.

\subsection{ Scaling of the Gap}

\begin{figure}[ht]
\includegraphics[width=0.5\textwidth]{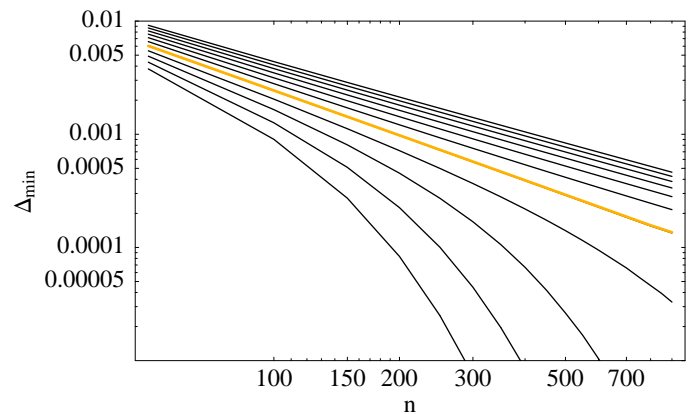}
\caption{\label{fig:slope} Scaling of the energy gap at the QPT point with $n$ for different values of the $\alpha$ parameter. The curves range from $\alpha = 0 $  to $\alpha = 3$. One observes a clear crossover at the critical value $\alpha_c=\frac{3 \sqrt{3}}{2}$ (yellow) with a $\nu$ value close to $\frac{4}{3}$.}	
\end{figure}

Fig.\ref{fig:slope} shows the scaling of the minimun energy gap, obtained along the path $H(s)$, with $n$.
For $\alpha<\frac{ 3\sqrt{3}}{2}$ the gap is proportional to $\frac{1}{n}$. This arises because of our normalization choice of the total Hamiltonian. Had we chosen a normalization in which the energy was a extensive quantity (which can be obtained by multiplying the total Hamiltonian by $n$) and the gap would be constant in the large $n$ limit. This result is simply derived from the standard Holstein-Primakoff method. At the second order transition point ($\alpha=\frac{ 3\sqrt{3}}{2}$) the numerical computation displays a clear slope crossover toward a still polynomial but non trivial exponent $\Delta \sim n^{-\nu} $ (Fig.\ref{fig:slope}). Numerically we find $\nu$ to be close to $\frac{4}{3}$, which is the value found previously in \cite{dusuel-2005-71}. for an also fully connected spin system but where the Hamiltonian was limited to two-body interactions. In the region where the first order QPT occurs $\alpha>\frac{ 3\sqrt{3}}{2}$ the gap vanishes exponentially with $n$. This is a general behavior for first order QPT.

\section{ Conclusion}

In this paper we discuss in detail an Hamiltonian evolution which should be viewed as a toy model for adiabatic computation. Indeed the phenomenological properties of this system correspond to what is usually expected in more realistic implementations: a Hamiltonian based on spin-spin interactions, a final Hamiltonian ($s=1$) diagonal in the computational basis and a non trivial behavior for some intermediate values of $s$ corresponding to a QPT. An interesting feature of our model is that it is build on a two dimensional parameter space which allows to trigger the Hamiltonian path form a trivial one (without QPT) to a regime where a first order QPT occurs, separated by a second order phase transition. The above phase space may serve as a template for more realistic cases. Given a classical computation problem the precise $H_P$ formulation and the choice of the initial Hamiltonian $H_0$ should result from an optimization process.
A qualitative knowledge of the phase space is required for that analysis (in terms of the topology of the first and second order phase transition manifolds).  In particular an heuristic point of view would lead to looking (in the phase space) for second order QPT to built an optimum Hamiltonian path.
\\
\\
\acknowledgments
It is a pleasure to thank Julien Vidal and Sebastien Dusuel for several interesting discussions in particular concerning the computation of finite size corrections. We also benefited from fruitful discussions with Remi Monasson about the subtle features of classically hard problems. PR was partially supported by FCT and EU FEDER through POCTI and POCI, namely via QuantLog POCI/MAT/55796/2004 Project of CLC-DM-IST, SQIG-IT and grant SFRH/BD/16182/2004/2ZB5.


\bibliography{biblio1}

\begin{thebibliography}{19}
\expandafter\ifx\csname natexlab\endcsname\relax\def\natexlab#1{#1}\fi
\expandafter\ifx\csname bibnamefont\endcsname\relax
  \def\bibnamefont#1{#1}\fi
\expandafter\ifx\csname bibfnamefont\endcsname\relax
  \def\bibfnamefont#1{#1}\fi
\expandafter\ifx\csname citenamefont\endcsname\relax
  \def\citenamefont#1{#1}\fi
\expandafter\ifx\csname url\endcsname\relax
  \def\url#1{\texttt{#1}}\fi
\expandafter\ifx\csname urlprefix\endcsname\relax\def\urlprefix{URL }\fi
\providecommand{\bibinfo}[2]{#2}
\providecommand{\eprint}[2][]{\url{#2}}

\bibitem[{\citenamefont{Farhi et~al.}(2000)\citenamefont{Farhi, Goldstone,
  Gutmann, and Sipser}}]{FarGolGutSip00}
\bibinfo{author}{\bibfnamefont{E.}~\bibnamefont{Farhi}},
  \bibinfo{author}{\bibfnamefont{J.}~\bibnamefont{Goldstone}},
  \bibinfo{author}{\bibfnamefont{S.}~\bibnamefont{Gutmann}}, \bibnamefont{and}
  \bibinfo{author}{\bibfnamefont{M.}~\bibnamefont{Sipser}}
  (\bibinfo{year}{2000}), \bibinfo{note}{quant-ph/0001106}.

\bibitem[{\citenamefont{Messiah}(1976)}]{Mes76}
\bibinfo{author}{\bibfnamefont{A.}~\bibnamefont{Messiah}},
  \bibinfo{journal}{Quantum Mechanics} \textbf{\bibinfo{volume}{Vol. II}},
  \bibinfo{pages}{Amsterdam: North Holland, New York: Wiley}
  (\bibinfo{year}{1976}).

\bibitem[{\citenamefont{van Dam et~al.}(2001)\citenamefont{van Dam, Mosca, and
  Vazirani}}]{DamMosVaz01}
\bibinfo{author}{\bibfnamefont{W.}~\bibnamefont{van Dam}},
  \bibinfo{author}{\bibfnamefont{M.}~\bibnamefont{Mosca}}, \bibnamefont{and}
  \bibinfo{author}{\bibfnamefont{U.}~\bibnamefont{Vazirani}},
  \bibinfo{journal}{Proc. 42nd FOCS} \textbf{\bibinfo{volume}{42}},
  \bibinfo{pages}{279} (\bibinfo{year}{2001}).

\bibitem[{\citenamefont{Aharonov et~al.}(2004)\citenamefont{Aharonov, van Dam,
  Kempe, Landau, Lloyd, and Regev}}]{AhaDamKemLanLloReg04}
\bibinfo{author}{\bibfnamefont{D.}~\bibnamefont{Aharonov}},
  \bibinfo{author}{\bibfnamefont{W.}~\bibnamefont{van Dam}},
  \bibinfo{author}{\bibfnamefont{J.}~\bibnamefont{Kempe}},
  \bibinfo{author}{\bibfnamefont{Z.}~\bibnamefont{Landau}},
  \bibinfo{author}{\bibfnamefont{S.}~\bibnamefont{Lloyd}}, \bibnamefont{and}
  \bibinfo{author}{\bibfnamefont{O.}~\bibnamefont{Regev}},
  \bibinfo{journal}{Proc. 45nd FOCS} p.~\bibinfo{pages}{42}
  (\bibinfo{year}{2004}).

\bibitem[{\citenamefont{Farhi et~al.}(2005)\citenamefont{Farhi, Goldstone,
  Gutmann, and Nagaj}}]{FarGolGutNag05}
\bibinfo{author}{\bibfnamefont{E.}~\bibnamefont{Farhi}},
  \bibinfo{author}{\bibfnamefont{J.}~\bibnamefont{Goldstone}},
  \bibinfo{author}{\bibfnamefont{S.}~\bibnamefont{Gutmann}}, \bibnamefont{and}
  \bibinfo{author}{\bibfnamefont{D.}~\bibnamefont{Nagaj}}
  (\bibinfo{year}{2005}), \bibinfo{note}{quant-ph/0512159}.

\bibitem[{\citenamefont{Reichardt}(2004)}]{Rei04}
\bibinfo{author}{\bibfnamefont{B.~W.} \bibnamefont{Reichardt}},
  \bibinfo{journal}{Proc. 36th STOC} p. \bibinfo{pages}{502}
  (\bibinfo{year}{2004}).

\bibitem[{\citenamefont{Wei and Ying}(2006)}]{wei-2006-}
\bibinfo{author}{\bibfnamefont{Z.}~\bibnamefont{Wei}} \bibnamefont{and}
  \bibinfo{author}{\bibfnamefont{M.}~\bibnamefont{Ying}}
  (\bibinfo{year}{2006}), \bibinfo{note}{quant-ph/0604077}.

\bibitem[{\citenamefont{Znidaric and Horvat}(2006)}]{ZniHor05}
\bibinfo{author}{\bibfnamefont{M.}~\bibnamefont{Znidaric}} \bibnamefont{and}
  \bibinfo{author}{\bibfnamefont{M.}~\bibnamefont{Horvat}},
  \bibinfo{journal}{Phys.Rev. A} \textbf{\bibinfo{volume}{73}},
  \bibinfo{pages}{022329} (\bibinfo{year}{2006}).

\bibitem[{\citenamefont{Farhi and Goldstone}(2000)}]{FarGol00}
\bibinfo{author}{\bibfnamefont{E.}~\bibnamefont{Farhi}} \bibnamefont{and}
  \bibinfo{author}{\bibfnamefont{J.}~\bibnamefont{Goldstone}}
  (\bibinfo{year}{2000}), \bibinfo{note}{quant-ph/0007071}.

\bibitem[{\citenamefont{Farhi et~al.}(2001)\citenamefont{Farhi, Goldstone,
  Gutmann, Lapan, Lundgren, and Preda}}]{FarGolGutLapLunPre01}
\bibinfo{author}{\bibfnamefont{E.}~\bibnamefont{Farhi}},
  \bibinfo{author}{\bibfnamefont{J.}~\bibnamefont{Goldstone}},
  \bibinfo{author}{\bibfnamefont{S.}~\bibnamefont{Gutmann}},
  \bibinfo{author}{\bibfnamefont{J.}~\bibnamefont{Lapan}},
  \bibinfo{author}{\bibfnamefont{A.}~\bibnamefont{Lundgren}}, \bibnamefont{and}
  \bibinfo{author}{\bibfnamefont{D.}~\bibnamefont{Preda}},
  \bibinfo{journal}{Science} \textbf{\bibinfo{volume}{292 (5516)}},
  \bibinfo{pages}{472} (\bibinfo{year}{2001}).

\bibitem[{\citenamefont{Roland and Cerf}(2003)}]{RolCer03}
\bibinfo{author}{\bibfnamefont{J.}~\bibnamefont{Roland}} \bibnamefont{and}
  \bibinfo{author}{\bibfnamefont{N.~J.} \bibnamefont{Cerf}},
  \bibinfo{journal}{Phys. Rev. A} \textbf{\bibinfo{volume}{68}},
  \bibinfo{pages}{062312} (\bibinfo{year}{2003}).

\bibitem[{\citenamefont{Andrew M.~Childs}(2002)}]{ChiFarPre02}
\bibinfo{author}{\bibfnamefont{J.~P.} \bibnamefont{Andrew M.~Childs},
  \bibfnamefont{Edward~Farhi}}, \bibinfo{journal}{Phys. Rev. A}
  \textbf{\bibinfo{volume}{65}}, \bibinfo{pages}{012322}
  (\bibinfo{year}{2002}).

\bibitem[{\citenamefont{Wootters}(1998)}]{wootters-1998-80}
\bibinfo{author}{\bibfnamefont{W.~K.} \bibnamefont{Wootters}},
  \bibinfo{journal}{Phys. Rev. Lett.} \textbf{\bibinfo{volume}{80}},
  \bibinfo{pages}{2245} (\bibinfo{year}{1998}).

\bibitem[{\citenamefont{Dusuel and Vidal}(2005)}]{dusuel-2005-71}
\bibinfo{author}{\bibfnamefont{S.}~\bibnamefont{Dusuel}} \bibnamefont{and}
  \bibinfo{author}{\bibfnamefont{J.}~\bibnamefont{Vidal}},
  \bibinfo{journal}{Phys. Rev. B} \textbf{\bibinfo{volume}{71}},
  \bibinfo{pages}{224420} (\bibinfo{year}{2005}).

\bibitem[{\citenamefont{Wang and Molmer}(2002)}]{wang-2002-18}
\bibinfo{author}{\bibfnamefont{X.}~\bibnamefont{Wang}} \bibnamefont{and}
  \bibinfo{author}{\bibfnamefont{K.}~\bibnamefont{Molmer}},
  \bibinfo{journal}{Eur. Phys. J. D} \textbf{\bibinfo{volume}{18}},
  \bibinfo{pages}{385} (\bibinfo{year}{2002}).

\bibitem[{\citenamefont{Vidal}(2006)}]{vidal-2006-}
\bibinfo{author}{\bibfnamefont{J.}~\bibnamefont{Vidal}},
  \bibinfo{journal}{Phys. Rev. A} \textbf{\bibinfo{volume}{73}},
  \bibinfo{pages}{062318} (\bibinfo{year}{2006}).

\bibitem[{\citenamefont{Vidal et~al.}(2004)\citenamefont{Vidal, Palacios, and
  Mosseri}}]{vidal-2004-69}
\bibinfo{author}{\bibfnamefont{J.}~\bibnamefont{Vidal}},
  \bibinfo{author}{\bibfnamefont{G.}~\bibnamefont{Palacios}}, \bibnamefont{and}
  \bibinfo{author}{\bibfnamefont{R.}~\bibnamefont{Mosseri}},
  \bibinfo{journal}{Phys. Rev. A} \textbf{\bibinfo{volume}{69}},
  \bibinfo{pages}{022107} (\bibinfo{year}{2004}).

\bibitem[{\citenamefont{Barthel et~al.}(2006)\citenamefont{Barthel, Dusuel, and
  Vidal}}]{barthel-2006-}
\bibinfo{author}{\bibfnamefont{T.}~\bibnamefont{Barthel}},
  \bibinfo{author}{\bibfnamefont{S.}~\bibnamefont{Dusuel}}, \bibnamefont{and}
  \bibinfo{author}{\bibfnamefont{J.}~\bibnamefont{Vidal}}
  (\bibinfo{year}{2006}), \bibinfo{note}{cond-mat/0606436}.

\bibitem[{\citenamefont{Ribeiro et~al.}()\citenamefont{Ribeiro, Vidal, and
  Mosseri}}]{pedro-2006-}
\bibinfo{author}{\bibfnamefont{P.}~\bibnamefont{Ribeiro}},
  \bibinfo{author}{\bibfnamefont{J.}~\bibnamefont{Vidal}}, \bibnamefont{and}
  \bibinfo{author}{\bibfnamefont{R.}~\bibnamefont{Mosseri}}, \bibinfo{note}{(in
  preparation)}.

\end{thebibliography}

\end{document}